\documentclass[twocolumn,aps,prl,amssymb,amstext,showpacs]{revtex4-1}

\usepackage{graphicx,amsfonts,amssymb,amsmath,bm}
\usepackage{color}
\usepackage{ulem}

\newcommand\half{\tfrac{1}{2}}

\newcommand\zhat{\bm{\hat z}}
\newcommand\xhat{\bm{\hat x}}
\newcommand\yhat{\bm{\hat y}}

\renewcommand\Im{\operatorname{Im}}

\begin{document}

\title{Dynamics of ferromagnetic domain walls under extreme fields}

\author{Arseni Goussev$^{1,2}$, JM Robbins$^3$, Valeriy Slastikov$^3$, Sergiy Vasylkevych$^{3,4}$}

\affiliation{
	$^1${\color{black}School of Mathematics and Physics, University of Portsmouth, Portsmouth PO1 3HF, United Kingdom}\\
	$^2$Department of Mathematics, Physics and Electrical Engineering, Northumbria University, Newcastle Upon Tyne NE1 8ST, United Kingdom\\
	$^3$School of Mathematics, University of Bristol, University Walk, Bristol BS8 1TW, United Kingdom\\
	$^4$ Institute of Meteorology, University of Hamburg, Grindelberg 7,  D-20144 Hamburg, Germany}

\date{\today}

\begin{abstract}
We report the existence of a new regime for domain wall motion in {\color{black} uniaxial and} near-uniaxial ferromagnetic nanowires, characterised by  applied magnetic fields
sufficiently strong that one of the domains becomes unstable. 
{\color{black} There appears a new stable solution of the Landau-Lifshitz-Gilbert equation, describing a nonplanar domain wall moving with constant  velocity  and precessing with constant frequency.  
Even in the presence of thermal noise, the new solution can propagate for distances on the order of 500 times the field-free domain wall width before  fluctuations in the unstable domain become appreciable.}
\end{abstract}

\maketitle

%

The dynamical response of magnetic domains in ferromagnetic nanostructures to applied fields and spin-polarized currents offers rich physics \cite{Tatara04, Beach05, hayashi:2006, Hayashi08, Thomas2010}, presents unresolved  mathematical challenges \cite{Braun12, Hellman17}, and promises exciting technological applications \cite{Allwood05, Parkin08}.
Of particular importance is the problem of domain wall motion, in which a ferromagnetic material has two neighbouring magnetic domains, one expanding and the other contracting under the action of an applied field.
To date, this problem has been addressed, analytically and numerically, in nanoscale systems with a variety of geometries and topologies, including  tubes, ribbons and films (see e.g. Refs.~\cite{Hertel2011, Goussev2014, Depassier2014, gaididei14, gaididei2017, boulle13}).  Here we focus on the important case of ferromagnetic nanowires \cite{Braun12, Tatara2008, ThiavilleBook06, goussev13}.

\begin{figure}[h]\begin{center}
  \includegraphics[width=0.36\textwidth]{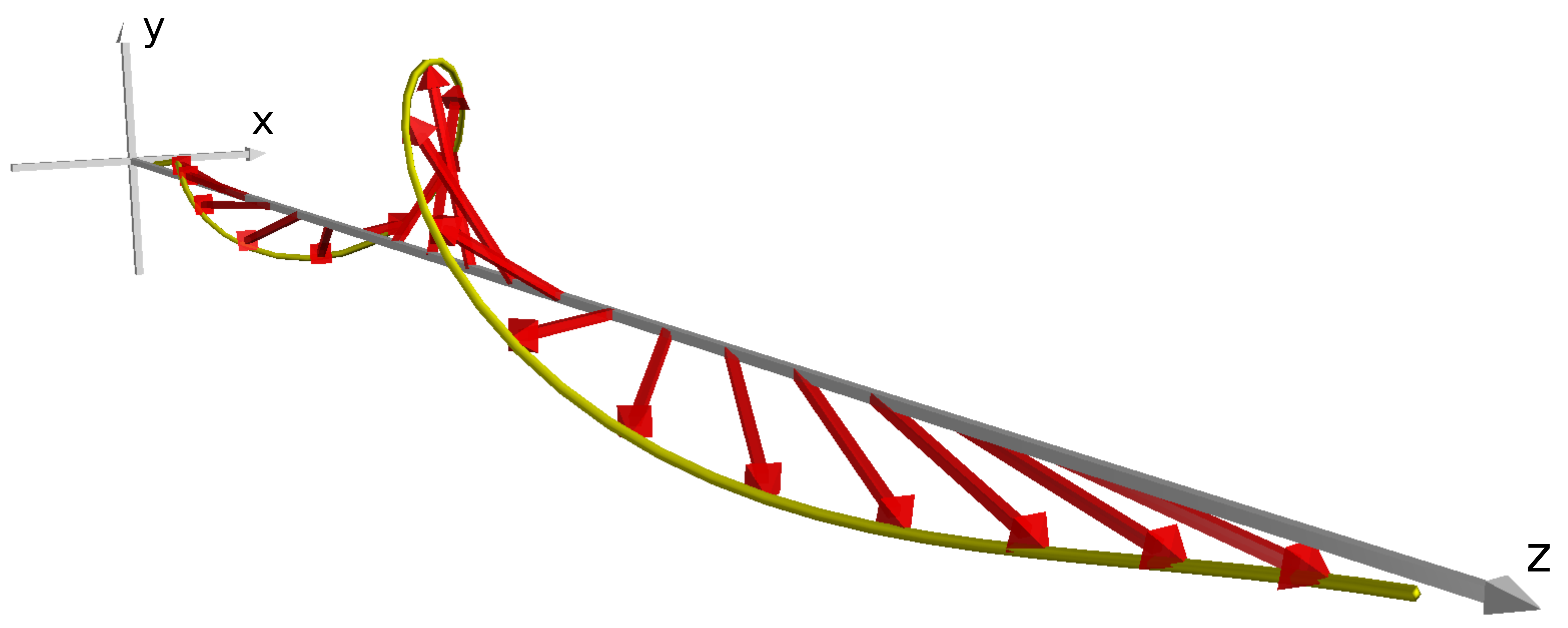}
\caption{High-field domain wall with tail-to-tail boundary conditions. {\color{black} The envelope (yellow curve) of the magnetisation (red arrows) indicates a helical as opposed to planar profile.  The asymptotic sense and pitch of the helix may be interpreted in terms of the chirality and wavelength of entrained spin waves.}}
\label{fig:arrow_profile}\end{center}
\end{figure}

A common feature of most of these studies (but cf Refs.~\cite{depassier15, benguria_depassier16}, discussed below) is the  assumption that the applied field is not strong enough to destabilise either domain. %
Here, we consider the case of applied fields sufficiently strong that one of the two domains becomes intrinsically unstable.
We show  that there emerges a 
fast-travelling  precessing domain wall with 
nonplanar profile -- see Fig.~\ref{fig:arrow_profile}, and calculate its velocity and precession frequency.
 We estimate the lifetime of the domain wall in the presence of thermal noise; for realistic parameters, it can travel  500 times the field-free domain-wall width before being 
 overtaken by thermal fluctuations.

We start from a standard model for domain wall dynamics under an applied field $H_a \bm{\hat z}$, taking the wire to be one dimensional along the $z$-axis.  For definiteness, we take $H_a > 0$.  The evolution of the magnetisation, $M_s \bm{m}(z,t)$, where $M_s$ is the fixed saturation magnetisation and the unit-vector $\bm{m} = (m_1,m_2,m_3)$ determines  orientation, is governed by the Landau-Lifshitz-Gilbert (LLG) equation,
\begin{equation}
\label{eq:LLG}
\partial_t \bm{m} = \gamma \bm{H} \times \bm{m} +\alpha \bm{m} \times\partial_t \bm{m} \, , 
\end{equation}
where $\bm{H} = -( M_s)^{-1} \delta E / \delta \bm{m} + H_a\bm{\hat z}$ is the effective magnetic field, $\gamma$ the gyromagnetic ratio and $\alpha$  the Gilbert damping constant (typically $\alpha \ll 1$).  The micromagnetic energy per unit cross-sectional area is given by
\begin{equation}
  E =  
\half  \int   \left( A \left|  \partial_z \bm{m} \right|^2
  + K (1- m_3^2) 
  +  K_2 m_2^2\right) dz,
\label{eq:energy}
\end{equation}
where $A$ is the exchange constant and $K, K_2 \ge 0$
are the anisotropy constants along the (easy) $z$- and (hard)
$y$-axes.  
The spatially uniform domains $\bm{m} = \pm \bm{\hat z}$ are  global minimisers of the energy,  so that  boundary conditions appropriate for a (head-to-head) domain wall are $\bm{m}(\pm\infty,\cdot) = \mp \bm{\hat z}$.  
This description incorporates several simplifications, including reducing to one dimension and  incorporating  the magnetostatic energy into the local anisotropy; see \cite{sanchez09, slastikov12} for discussion and justification.
%
%
%
%
%
%


The model \eqref{eq:LLG}--\eqref{eq:energy} has been extensively analysed in the literature (see e.g.~\cite{Walker74, slonczewski79, kosevich90, ThiavilleBook06, Hertel2010, GRS10Domain, goussev13, slastikov19}).
We will restrict our attention to the case of near-uniaxial wires, for which $K\gg K_2$ {\color{black}(eventually, we will take $K_2 = 0$)}.  For applied fields $H_a$ below the Walker breakdown field $H_W = \alpha K_2/(2M_s) $, there appears an explicit  stable travelling wave solution, $\bm{m_*}(z-vt)$,  with  velocity depending nonlinearly on $H_a$; for $H_a = H_W$, the Walker breakdown  velocity is  $V_W 
= \frac{\gamma}{  M_s} \sqrt{\frac{A}{4K + 2K_2}} K_2$ \cite{Walker74}.    The Walker profile $\bm{m_*}$ lies in a fixed plane whose inclination to the $x$-axis increases with $H_a$ up to a maximum of $45^\circ$ at breakdown.  

For fields above breakdown, 
the dynamics is more complicated.  While there is no known explicit solution, numerical simulations, collective coordinate models and  asymptotic analysis reveal profiles in which the magnetisation is no longer planar and  executes periodic motion, including translation, precession 
and breathing (see e.g.~\cite{ThiavilleBook06, goussev13}). The mean velocity of the domain wall  actually decreases with increasing $H_a$.
For  large enough applied fields so that $K_2$ can be neglected (but still with both domains stable), the behaviour approaches a simple explicit solution in which 
the static planar {\color{black} uniaxial}  
profile 
moves with uniform velocity $V_p = \alpha \gamma H_a / M_s \color{black} \ll V_W$ 
and precession frequency $\Omega_p = \gamma H_a$ \cite{GRS10Domain}.
 
The preceding description of domain wall dynamics  applies when 
 the spatially uniform domains $\bm{m} = \pm\bm{\hat z}$ 
are  energetically stable; the condition for  stability is  $|H_a| < K/M_s$.  For $H_a >K/M_s$, the uniform domain $\bm{m} =- \bm{\hat z}$ becomes unstable, and under  perturbations, e.g.~thermal fluctuations,  switches
 spontaneously to  $+\bm{\hat z}$.  
 
\begin{figure}[h]
\begin{center}
  \includegraphics[width=0.28\textwidth]{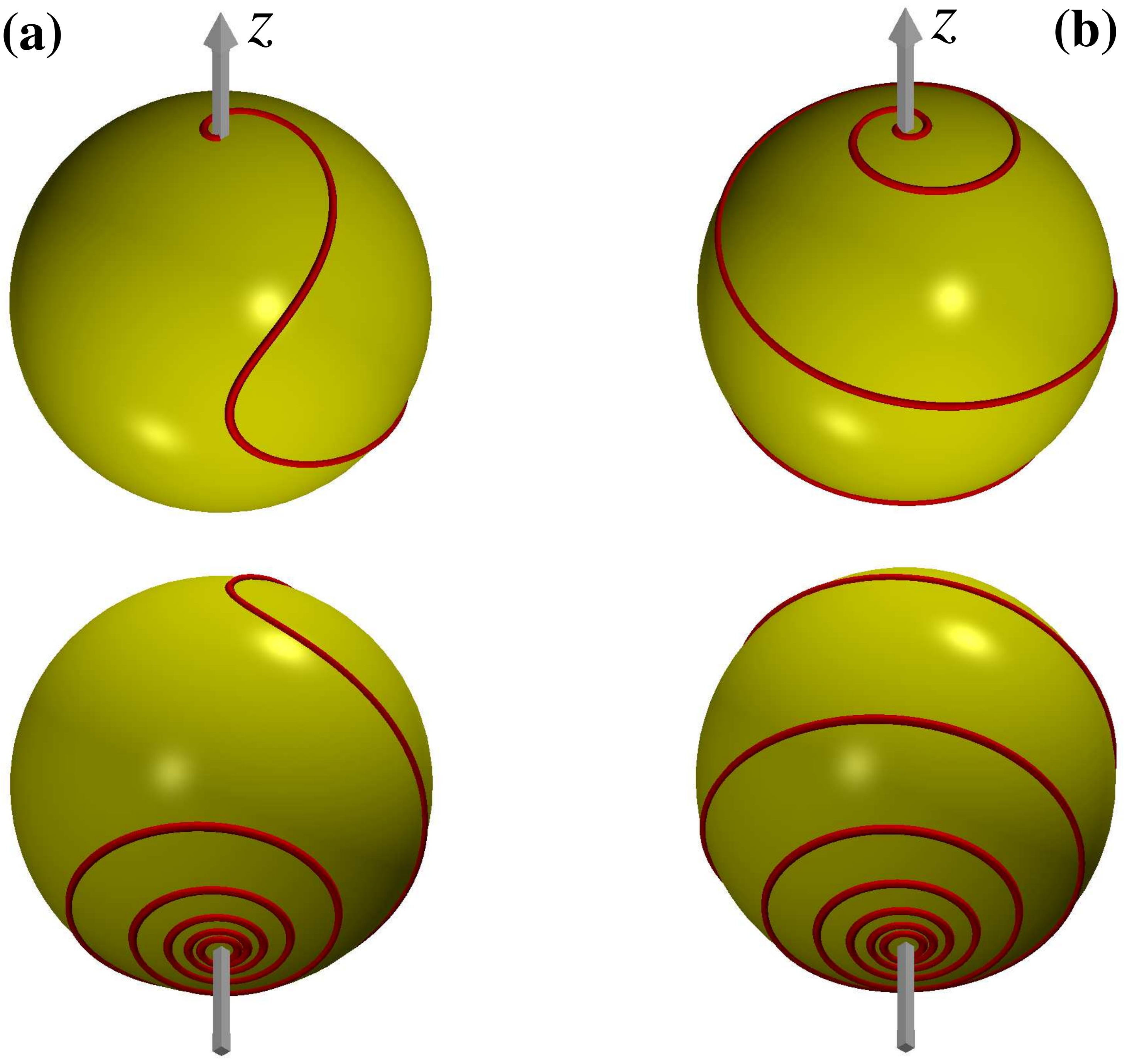}
\caption{Two spherical pendulum trajectories, shown from  perspectives above and below the sphere.  
\color{black}
In (a), with $h_a = 2.3$, the sense of the azimuthal rotation changes sign as the trajectory passes from the north to the south pole. In (b), with $h_a = 5$,  the sense of rotation stays the same.
 In both cases, 
$\alpha = 0.1$, and $v$ and $\omega$ are given by Eq.~\eqref{eq: v and omega}.}
\label{fig: orbit on sphere}\end{center}
\end{figure}

 A similar switching process takes place in the unstable tail of a domain wall.   
 However, as we  report here, before this occurs, there emerges a new, persistent domain-wall dynamics distinct from the well-known behaviour for $H_a < K/M_s$.   The high-field profile is strongly nonplanar; the tails are helical with pitches that may have the same or opposite signs -- see Fig.~\ref{fig: orbit on sphere}.  
 The velocity of the high-field domain wall scales nonlinearly with applied field, and
for suitable parameters  is comparable to or may substantially exceed the Walker breakdown velocity for strongly anisotropic nanowires.

To simplify the analysis, we consider the strictly uniaxial case $K_2 = 0$, so that the problem has rotational symmetry about $\zhat$; it turns out that the behaviour for small, nonzero $K_2$ is qualitatively similar.  
It is also convenient to introduce dimensionless variables  $\zeta = \sqrt{K/A} \, z$ and $\tau = (\gamma K/M_s)\, t$.
{\color{black} Then the LLG equation~\eqref{eq:LLG} becomes
\begin{equation}\label{eq: LLG ND}
\bm{\dot m} = \left( \bm{m''}  + m_3 \bm{\hat z} + h_a \bm{\hat z}\right) \times \bm{m} + \alpha \bm{m} \times \bm{\dot m},
\end{equation}
in which the only  (dimensionless) parameters are $\alpha$ and $h_a = (M_s/K) H_a$, the rescaled applied field.  In these units,   the static (field-free) domain wall has unit width.}


We look for solutions {\color{black} of 
 Eq.~\eqref{eq: LLG ND} travelling  with fixed (dimensionless) 
 velocity $v$ 
 and precessing  with fixed (dimensionless) frequency $\omega$.  These are of the form
\begin{equation}\label{eq: ansatz}
\bm{m}(\zeta,\tau) = {\mathcal R}_3(\omega \tau ) \bm{n}(\zeta - v{\color{black} \tau}),
\end{equation}
where ${\mathcal R}_3(\phi)$ denotes the rotation about $\zhat$ by angle $\phi$, and $\bm{n}$ is the domain wall profile.}
 Substituting   \eqref{eq: ansatz} into 
  \eqref{eq: LLG ND}, we get the following second-order  ODE for $\bm{n}$: 
 \begin{equation}
\label{eq: ODE n}
\bm{n''} = (\omega - n_3 - h_a) \zhat - v \bm{n} \times \bm{n'} + \alpha(\omega  \zhat \times \bm{n}- v \bm{n'})  - \lambda \bm{n},
\end{equation}
where 
$\lambda = |\bm{n'}|^2  - (n_3 + h_a-\omega) n_3  $.

{\color{black} While the ODE \eqref{eq: ODE n} cannot be solved explicitly, we can obtain the main qualitative features of the high-field 
profile through a dynamical-systems analysis.  To this end, it is helpful to introduce the following mechanical analogy.  
We temporarily regard $\bm{n}(\zeta)$ 
as the position of a particle moving on the surface of a sphere, with $\zeta$ regarded  as a fictitious time coordinate.  From this point of view, \eqref{eq: ODE n} describes the dynamics of a spherical pendulum (of unit length,  mass and  charge)  subject to a uniform gravitational force $-(h_a - \omega) \zhat$ as well as the following additional forces:  
(i) a Lorentz force, $v \bm{n} \times \bm{n'}$, arising from a radial magnetic field of uniform strength (
which may be interpreted as 
the field of a magnetic monopole of charge $-v$ at the centre of the sphere); (ii) a harmonic force arising from a potential  $\tfrac{1}{2} n_3^2$;  (iii) a 
damping force, $-\alpha v \bm{n'}$; and (iv) a nonconservative azimuthal torque, $\alpha\omega  \zhat \times \bm{n}$.  Finally, there is (v)  a force of constraint, $\lambda \bm{n}$, ensuring that the length of the pendulum remains fixed. 
We remark that for $\alpha = 0$, Eq.~\eqref{eq: ODE n}, regarded as a Hamiltonian system, is integrable}, with energy ${\mathcal E} = \half \bm{n'}^2 + (\half n_3 + h_a-\omega) n_3 $ and canonical angular momentum
 ${\mathcal L} = \zhat\cdot(\bm{n}\times \bm{n'}) - v n_3$ as conserved quantities.
 
The dynamics is no longer exactly solvable for $\alpha > 0$.
However, 
 it is easy to establish that Eq.~\eqref{eq: ODE n} has just two equilibria, namely $\bm{n} = \sigma \zhat$, corresponding to the pendulum  at rest and either upright ($\sigma = +1$) or downright ($\sigma = -1$).  In fact, we are seeking a trajectory which connects these two equilibria - a heteroclinic orbit $\bm{n}(\zeta)$ - with the pendulum upright at $\zeta =-\infty$ and downright at $\zeta = +\infty$; 
 this corresponds to a domain wall profile with the specified boundary conditions.  
 
 {\color{black} In order for such a heteroclinic orbit 
 to exist
for a range of values of $v$ and $\omega$, it turns out that we must require $+\zhat$ to be a saddle point and $-\zhat$ to be a stable node.  
To determine when these  conditions hold, we consider the linearised dynamics about the two equilibria.
For convenience, we write $\bm{n} = \sigma (\zhat + \epsilon (\eta_1 \xhat + \eta_2 \yhat)) + O(\epsilon^2)$ and introduce the complex coordinate $\eta = \eta_1 + i\eta_2$.  Substituting into Eq.~\eqref{eq: ODE n}}, we obtain the linearised equation 
 \begin{equation}\label{eq: linearised eta}
 \eta'' + rv \eta' - (1 + \sigma h_a + i r \omega) \eta = 0,  \end{equation}
 where $r = \alpha + i\sigma$.
 The associated characteristic equation (obtained by substituting $\eta = e^{i k \zeta}$) is \footnote{We note that if $\eta$ satisfies \eqref{eq: linearised eta}, then so does {\color{black} $e^{i\beta} \eta$ for any fixed $\beta$} (a consequence of azimuthal symmetry). Thus, $\eta$ and $i\eta$ correspond to independent solutions of 
 \eqref{eq: linearised eta}.} 
 \begin{equation}\label{eq: char equation}
  k^2 -i rvk + (1 + \sigma h_a + ir \omega) = 0.
  \end{equation}

{\color{black} The stabilities of $\sigma \zhat$ 
 are determined by the imaginary parts 
of the roots $k_\pm$ of 
\eqref{eq: char equation}.
For $\sigma = 1$, it is straightforward to establish that   $\Im k_\pm$ 
have opposite signs provided $h_a > 1$, in which case
$+\zhat$ is a saddle point for all $v$ and $\omega$. 
For $\sigma = -1$, 
 it is straightforward to establish that i)
$\Im k_\pm$ 
 have the same sign provided $ \omega^2 < (h_a - 1) v^2$, in which case $-\zhat$ is a node, and  ii) $-\zhat$ is a stable node provided $v>0$.
%
Thus, the conditions for the existence of a heteroclinic orbit over a range of values of $v$ and $\omega$ are}
 \begin{equation}\label{eq: -z stability threshold}
 v > 0  \text{ and }  \omega^2 < (h_a - 1) v^2. 
  \end{equation}
 
The heteroclinic orbit $\bm{n}(\zeta)$ 
 is unique up to 
  rotation about the $\zhat$-axis 
  and translation in $\zeta$.
 Via  
 Eq.~\eqref{eq: ansatz}, it corresponds  to a 
 travelling-wave solution of the LLG equation with velocity $v$ and precession frequency $\omega$. 
  %
Numerical solution of Eq.~\eqref{eq: ODE n} confirms the existence of this heteroclinic  orbit  
when Eq.~\eqref{eq: -z stability threshold} is satisfied; 
%
representative examples 
are shown in Fig.~\ref{fig: orbit on sphere}  \footnote{We remark that when Eq.~\eqref{eq: -z stability threshold} is violated by increasing $\omega^2$ above $(h_a - 1) v^2$,  
 the system undergoes a 
Hopf bifurcation. $-\zhat$ becomes a saddle, and a limit cycle appears 
on the line of latitude ${\color{black} n_3}  = -h_a v^2/(v^2+\omega^2)$ with precession frequency $\Omega = \omega/v$.}.

\begin{figure}[h]
\begin{center}
  \includegraphics[width=0.46\textwidth]{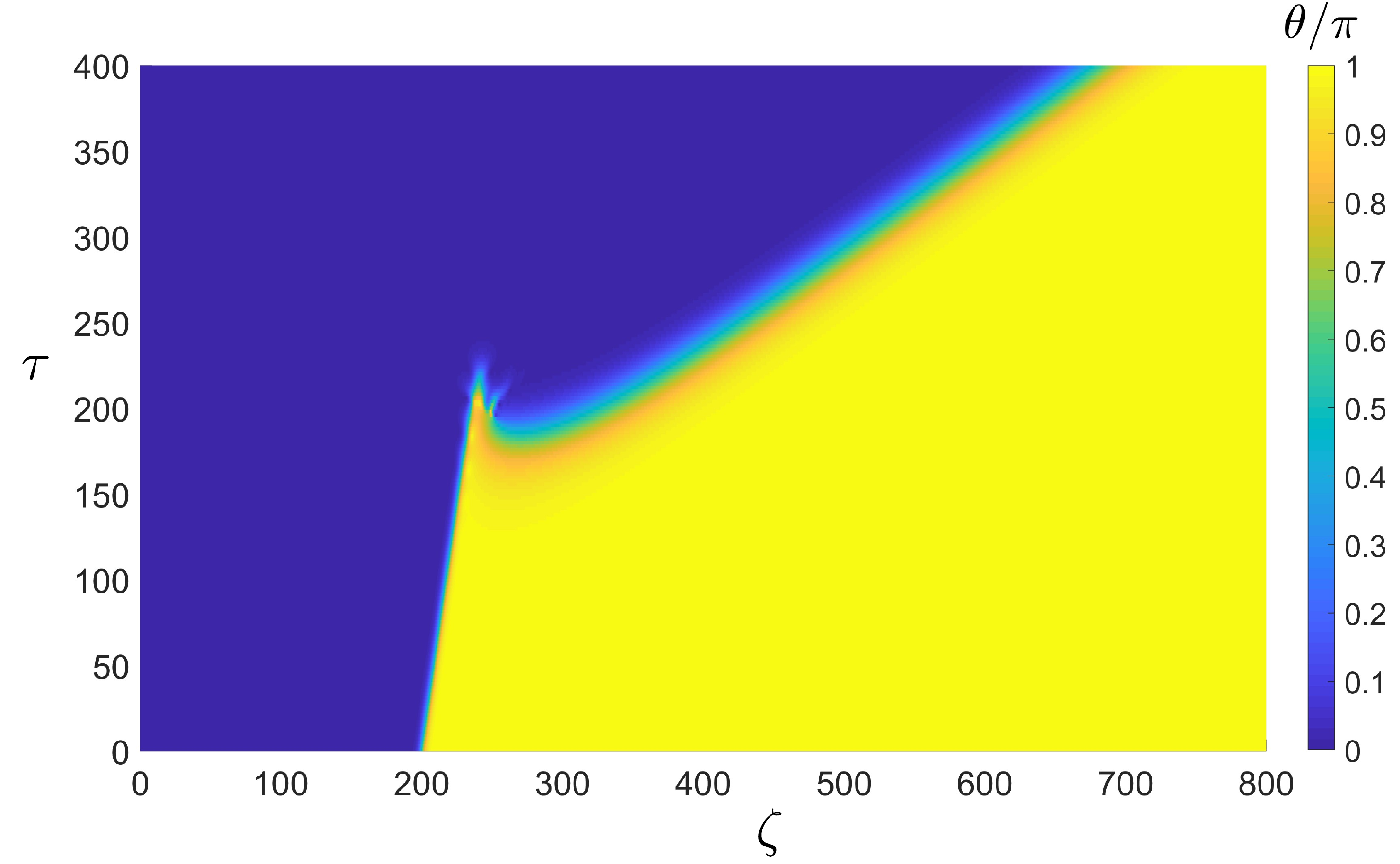}
\caption{Emergence of high-field profile: the 
evolution 
of the polar angle $\theta(\zeta,\tau) = \cos^{-1}(m_3)$ obtained from numerical solution of the LLG equation~\eqref{eq: LLG ND} with  static (field-free) domain wall profile as initial condition.  \color{black} Here $h_a = 3$ and $\alpha = 0.1$.}
\label{fig: onset of stable solution from initial condition}
\end{center}
\end{figure}

Numerical solution of the LLG equation~\eqref{eq: LLG ND} reveals the following surprising behaviour: 
For  initial conditions describing a sufficiently sharp head-to-head domain wall, the evolving profile approaches a 
traveling wave solution Eq.~\eqref{eq: ansatz} with  {\it specific}  values of $v$ and $\omega$.  
The selected velocity and precession frequency depend only on  $h_a$ and $\alpha$, and not on the initial condition.  This is illustrated in 
Fig.~\ref{fig: onset of stable solution from initial condition}, where 
the initial configuration is taken to be the static (field-free) domain wall profile.  At first, the evolution follows the exact precessing solution \cite{GRS10Domain}. 
The precessing solution is unstable, {\color{black} however \cite{Gou11}}, and after a short time, the new high-field 
profile emerges, with  much higher velocity. 

%

For scalar PDEs, there is a well-established method for determining the selected velocity of travelling-wave solutions based on the theory of front propagation into unstable states (see, eg, \cite{saarloos03} and references therein).  Here, we adapt this method for the vector-valued LLG equation~\eqref{eq: LLG ND}. The idea is to linearise the LLG equation 
 in the region of the unstable  tail of the profile, 
 {\color{black} ie where $\zeta \gg 1$, and  find a frame of reference in which, at long times, the propagating solution is nearly stationary.  With
 \begin{equation*}
 \bm{m} = -(\zhat + i\epsilon (\eta_1 \xhat + \eta_2\yhat)) + O(\epsilon^2), \quad \eta = \eta_1 + i\eta_2,
 \end{equation*}
  the linearised LLG equation for $\eta(\zeta,\tau)$ is given by 
 \begin{equation}\label{eq: linearised LLG}
 (1+i\alpha) \dot  \eta = i \eta'' + i(h_a - 1) \eta. 
 \end{equation}
The  solution is given explicitly by
\begin{gather} \label{eq: eta solution} \eta(\zeta,\tau) = \int \hat\eta_0(k) e^{i(k\zeta - \Omega(k) \tau)} \,dk, 
\text{ where}\\
 \label{eq: Omega(k)} \Omega(k) = -(h_a - 1 - k^2)/(1+i\alpha). 
 \end{gather}
In a frame moving with velocity $v$ and precessing with frequency $\omega$, the profile appears as $\tilde \eta(\zeta,\tau) = \color{black} e^{-i\omega \tau}  \eta(\zeta - v\tau, \tau)$, with integral representation 
\begin{equation}\label{eq: etatilde solution} \tilde\eta(\zeta,\tau) =  \int \hat\eta_0(k) e^{ i(kv -  \Omega(k) - \omega) \tau} \,  e^{ik\zeta}\, dk.\end{equation}
For long times $\color{black} \tau$,  the integral in \eqref{eq: etatilde solution} 
may be evaluated by the method of steepest descent; the contour is deformed through the (complex) saddle point $k_*$, characterised by 
\begin{equation}\label{eq: saddle point} \Omega'(k_*) = v,  \ \ \Im k_* > 0. \end{equation}
Evaluation of \eqref{eq: etatilde solution} yields 
\begin{equation} \tilde\eta(\zeta,\tau) \approx \frac{ \hat \eta_0(k_*)}{(2\pi \Omega''(k_*) \tau)^{1/2}}\,  e^{i(k_*v -  \Omega(k_*) - \omega) \tau} \, e^{ik_* \zeta}. \end{equation}
We choose $v$ and $\omega$ so that $\tilde\eta(\zeta,\tau)$ is $\tau$-independent (apart for a diffusive prefactor $\tau^{-1/2}$), ie so that
\begin{equation} \label{eq: no tau dependence} k_*v  = \Omega(k_*) - \omega.\end{equation}
With some calculation, Eqs.~\eqref{eq: Omega(k)}, \eqref{eq: saddle point} and \eqref{eq: no tau dependence} yield
%
\begin{equation}
\label{eq: v and omega}
v = 2\left(\frac{h_a-1}{1+\alpha^2}\right)^{1/2}, \quad \omega = 2\, \frac{h_a-1}{1+\alpha^2}.
\end{equation}
We note that it is precisely when $v$ and $\omega$ are given by \eqref{eq: v and omega} that  the roots of \eqref{eq: char equation} with $\sigma = -1$ coincide. 
This phenemenon is well known for scalar PDEs {\color{black} of reaction-diffusion type}, for example the KPP equation \cite{KPP}. 
}

\begin{figure}[h]
\begin{center}
  \includegraphics[width=0.48\textwidth]{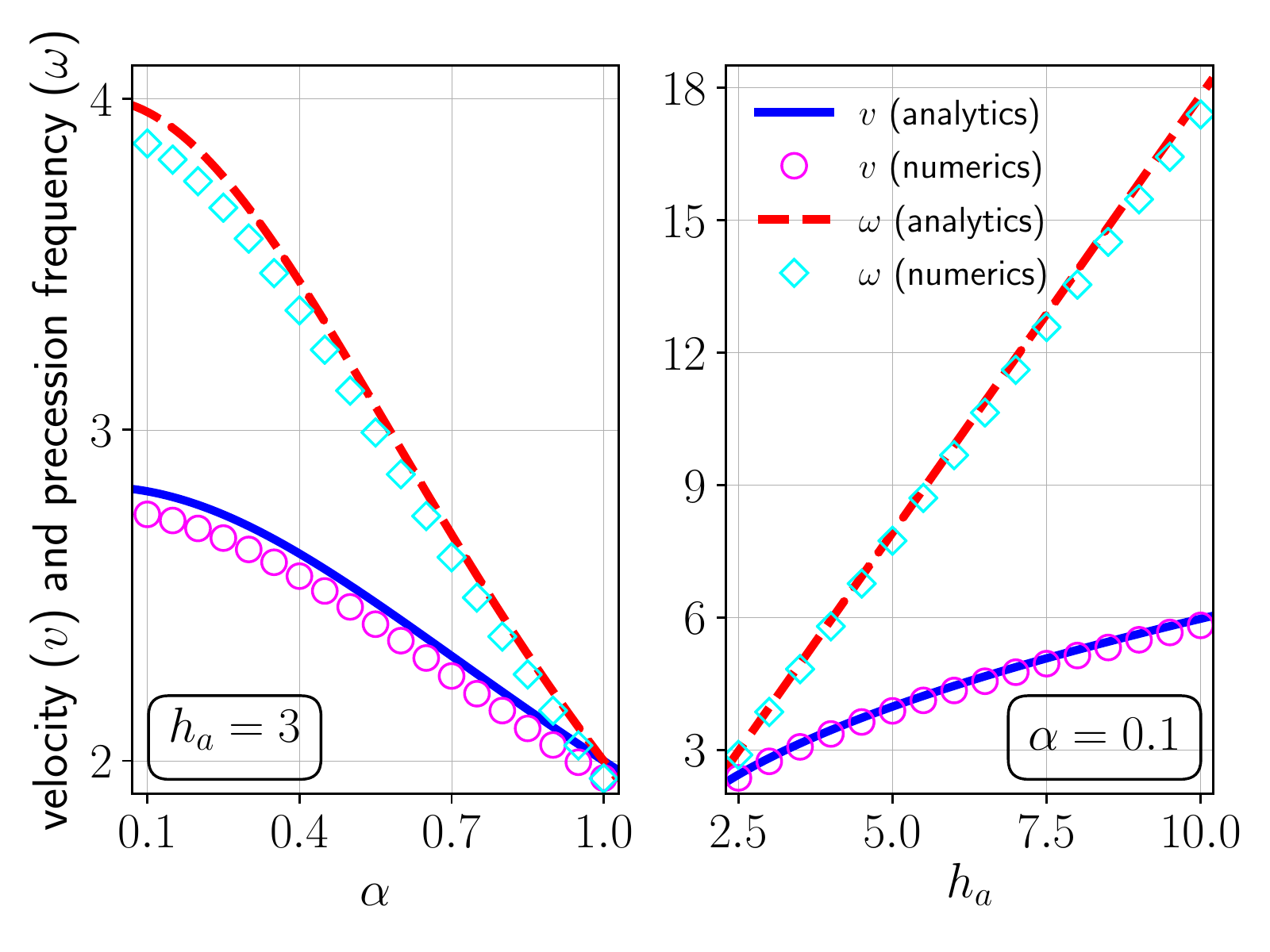}
\caption{Velocity (circles) and precession frequency (diamonds) of the  
high-field profile obtained from numerical solution of the LLG equation~\eqref{eq: LLG ND},  along with the analytic predictions of Eq.~\eqref{eq: v and omega}, plotted as functions of (a)  damping constant $\alpha$ for $h_a = 3$ and (b) applied field $h_a$ for $\alpha = 0.1$.}
\label{fig: velocity and frequency vs alpha}
\end{center}
\end{figure}

Confirmation of the preceding theory %
is provided in Fig.~\ref{fig: velocity and frequency vs alpha}.  {\color{black} We solve the LLG equation~\eqref{eq: LLG ND} numerically for a variety of initial conditions, using a finite difference scheme on a uniform rectangular grid, where spatial derivatives are represented by central finite differences with Neumann boundary conditions.   A time step is calculated via an explicit fourth-order Runge-Kutta method. In order to exactly maintain the constraint on the magnetization norm, the solution is renormalized after each time step.}
We determine the (initial-condition-independent) velocity and precession frequency of the emergent profile as functions of $h_a$ and of $\alpha$.  These are in good agreement with the analytic formulas \eqref{eq: v and omega}.
Numerically computed  profiles are shown in the Supplemental Material \footnote{See Supplemental Material at [URL will be inserted by publisher]
for a comparison of a numerically computed high-field domain wall
profile with analytic results.}.  They coincide with solutions of the ODE \eqref{eq: ODE n} with $v$ and $\omega$ given by \eqref{eq: v and omega}.  In particular, the chiralities of the domain wall tails are obtained from \eqref{eq: char equation}.

As noted previously, with $h_a > 1$, {\color{black}  the uniform profile $\bm{m} = -\zhat$ is unstable. 
It follows that the high-field profile is unstable to perturbations in the region $\zeta \gg 1$, for example due to thermal excitation of spin waves. To  estimate the time scale for this instability to set in,  we model this region as a cylindrical nanowire of  
finite length $L \gg \delta_{ex}$, where the exchange length, $\delta_{ex} =  \sqrt{A/K}$, is the width of the field-free domain wall.   (The estimate turns out to be independent of the choice of $L$.) The magnetisation 
is governed by the linearised LLG equation \eqref{eq: linearised LLG} 
with  transverse component, $\eta(\zeta,\tau)$, given by \eqref{eq: eta solution} but with the $k$-integral replaced by a sum over spin wave modes of wavenumber $k_j$,   with spin wave amplitudes   $\hat\eta_0(k_j)$ and (complex) frequencies $\Omega(k_j)$.    The
phases $\arg \hat\eta_0(k_j)$ are uncorrelated, so that the mean squared amplitude $|\eta|^2$ is the sum of the squared amplitudes of the spin waves.  
We suppose the magnetic field is applied from $\tau = 0$ onwards,  and let $\tau_c$ denote the time required for 
$|\eta|^2$ to equal one.  

As a crude approximation, we suppose that  only  spin waves with wavelengths greater than $\delta_{ex}$ contribute; the number of such spin waves is approximately $L/\delta_{ex}$.  Moreover, for these spin waves, we replace  $|\hat\eta(k_j)|$ and $\Omega(k_j)$ by their long wavelength limits $|\hat\eta_0|$ and 
$\Omega_0$, 
replacing $k_j$ by $k_0 = 1/L$ (more careful calculation does not change the estimate appreciably).  We obtain $|\eta(\zeta,\tau)|^2 \approx (L/\delta_{ex})  |\hat\eta_0|^2 e^{2\Im \Omega_0 \tau}$, so that 
$2 \Im \Omega_0 \tau_c  \approx \log ((\delta_{ex}/L)/|\hat\eta_0|^2)$.  
After time $\tau_c$, the domain wall travels a distance (in units of the exchange length) 
\begin{equation}
\label{eq:  distance}
d_c = v\tau_c   
= \frac{1}{\alpha} \sqrt{\frac{1+\alpha^2}{h_a - 1}} \log\frac{\delta_{ex}/L}{|\hat\eta_0|^2},
\end{equation}
where $v$ is given by \eqref{eq: v and omega} and we have used \eqref{eq: Omega(k)} for $\Omega_0$.

The initial amplitude $|\hat \eta_0|$  may be estimated from a simple equipartition argument.  The associated spin wave energy is approximately  $|\hat \eta_0|^2 K SL$, where $S$ is the cross-sectional area of the wire (for long wavelengths, the exchange energy is negligible).  At temperature $T$, before the magnetic field is applied,  each spin wave mode has energy $k_B T$, where $k_B$ is Boltzmann's constant.  Thus,  
\begin{equation} |\hat\eta_0|^{2} = k_B T/(KSL).\end{equation}
To estimate $d_c$, we take as representative values $A = 10^{-11}$ J/m,  $M_s H_a = 2 K = 10^6$ J/m$^3$, $S = 100$ nm$^{2}$, $T = 100$K and $\alpha = 0.01$.  {\color{black} (For $M_s = 10^6 {\rm A/m}$, this corresponds to an applied field strength of 1 Tesla.)}   In this case, the high-field domain wall 
propagates for approximately 500 static domain-wall widths  before being overtaken by thermal instabilities. } 

{\color{black} It is interesting to compare the (unscaled) high-field domain wall velocity $V$ {\color{black} in a uniaxial wire with easy-axis anistropy $K$} 
to the   Walker velocity $V_W$ for a strongly anisotropic wire  with  easy-axis anisotropy $K$ and 
hard-axis anisotropy $K_2 > K$, 
For {\color{black} large applied field 
 in the uniaxial case and large $K_2$ in the 
 anisotropic case (and weak damping for both)}, 
\begin{equation} V/V_W \sim \sqrt{8 M_s H_a/K_2}.\end{equation}
Thus, 
 for    
 $H_a$   comparable to 
 $K_2/M_s$, the high-field domain wall velocity {\color{black} in the uniaxial wire is greater than the 
 Walker velocity in the 
 anisotropic wire.}

We have concentrated on the case of 
 uniaxial nanowires.  Numerical calculations reveal qualitatively similar behaviour for small nonvanishing hard-axis anisotropy -- ie, a new high-field domain wall profile with characteristic velocity and precession frequency. A perturbative analysis can be developed for small $K_2 > 0$.  

The dynamics of domain walls in 
nanowires under small applied fields and currents has been extensively studied.  Here we consider the response of a domain wall to an applied magnetic field strong enough to make one of the domains unstable.  Naively {\color{black} one} might imagine the unstable domain to reorient itself spontaneously and incoherently.  Surprisingly, we show that for small transverse anisotropy,  
there emerges a coherent reorientation, whereby
the energetically stable domain grows via the propagation of a travelling and precessing domain wall.  

{\color{black} The threshold for the high-field regime is $H_a > K/M_s$.  For an isotropic material such as permalloy, $K \simeq \tfrac{1}{4} \mu_0 M_s^2$ \cite{Muratov17}.  In particular, for permalloy, $M_s \simeq 800\,$kA/m \cite{Bain2001},  so that the threshold is given approximately by $\tfrac{1}{4} \mu_0 M_s \simeq 0.25\,$T.}}
We note that early experiments on domain-wall motion in iron-garnet films at applied fields above the anisotropy threshold  \cite{Logginov82, Ivanov83} indicate a sublinear velocity response 
compatible with  \eqref{eq: v and omega}.  Radiation damping at high fields is discussed in a related theoretical work \cite{Baryakhtar82}.

The high-field domain wall profile has novel features.  Unlike the well-known Walker profile, it is nonplanar 
with 
asymmetrical tails comprised of 
 spin-wave trains of different characteristic wavenumbers and helicities. 
The coherent magnetization switching is eventually overtaken by 
thermal fluctuations far into the unstable domain, but can persist over length scales of many hundreds of widths of the domain wall.   For realistic parameters, the domain wall velocity in the high-field regime can be comparable to or larger than the Walker velocity.

 Benguria and Depassier \cite{depassier15, benguria_depassier16} consider the complementary case of strong biaxial anisotropy $K \ll K_2$, characteristic of thin ferromagnetic films. There appear transitions (depending on $\alpha$ and $K/K_2$) between the Walker solution with velocity $v \sim H_a$ and a KPP-type solution (for which one of the domains is necessarily unstable) with $v \sim \sqrt{H_a}$.  In this regime, the magnetisation is confined to a plane, and the LLG equation reduces to a scalar equation of reaction-diffusion type, for which the theory of unstable front propagation is highly developed (see e.g. \cite{saarloos03}).   For the case of near-uniaxial wires  considered here, the LLG equation is a vectorial equation; much less is known about unstable front propagation for systems {\color{black} as opposed to scalar equations.}

We are grateful to L.P.~Ivanov for drawing our attention to  References \cite{Logginov82}--\cite{Baryakhtar82} and for interesting comments.  AG   thanks   EPSRC   for   support   under   grant
EP/K024116/1.    {\color{black} JMR, VS  and SV thank  EPSRC   for   support
under grant EP/K02390X/1.  JMR and VS thank the Isaac Newton Institute for Mathematical Sciences for support and hospitality during the programme {\it Mathematical Design of New Materials},  supported by EPSRC grant number EP/R014604/1.  JMR acknowledges support from a Lady Davis Visiting Professorship at Hebrew University and a University Research Fellowship from the University of Bristol.  }
 

 \bibliography{high_field.bib}
 

%
%

\end{document}


\title{Dynamics of ferromagnetic domain walls under extreme fields\\[0.2cm] {\Large Supplemental Material}}

\author{Arseni Goussev$^{1,2}$, JM Robbins$^3$, Valeriy Slastikov$^3$, Sergiy Vasylkevych$^{3,4}$}

\affiliation{
	$^1${\color{black}School of Mathematics and Physics, University of Portsmouth, Portsmouth PO
		1 3HF, United Kingdom}\\
	$^2$Department of Mathematics, Physics and Electrical Engineering, Northumbria University, N
	ewcastle Upon Tyne NE1 8ST, United Kingdom\\
	$^3$School of Mathematics, University of Bristol, University Walk, Bristol BS8 1TW, United K
	ingdom\\
	$^4$ Institute of Meteorology, University of Hamburg, Grindelberg 7,  D-20144 Hamburg, Germany}

\date{\today}

\maketitle

\section{Numerically computed profile of a high-filed domain wall}

\begin{figure}[h!]
	\includegraphics[width=0.5\textwidth]{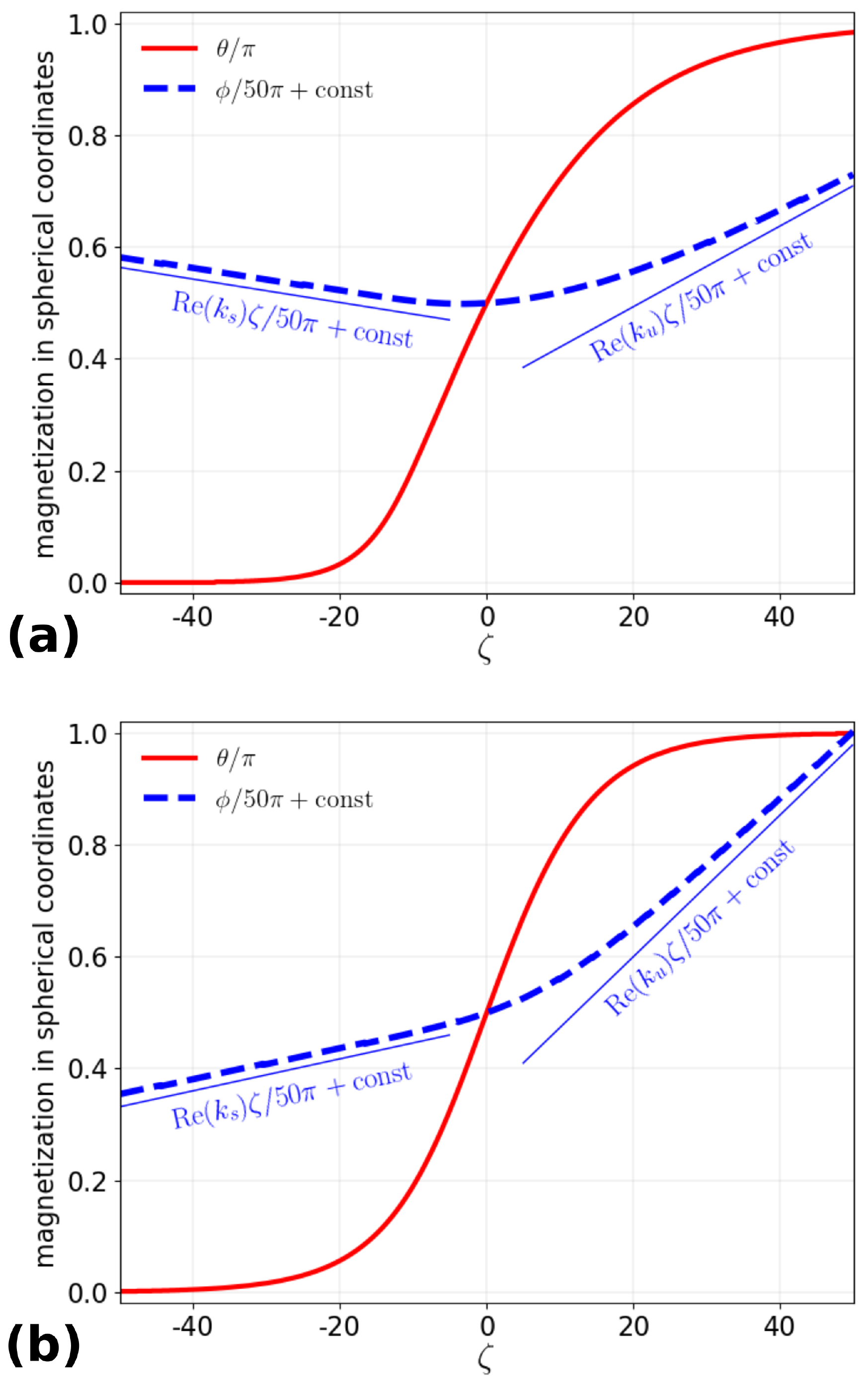}
	\caption{Polar coordinates $\theta$ and $\phi$ of the emergent high-field profile along with asymptotic wave numbers (the real parts of $k_{s,u}$ in Eq.~\eqref{eq: kappa_us}), which determine the rate of twisting in $\phi$ in the tails.    In (a) (cf Fig.~2a), the tails have opposite chirality, while in (b) (cf Fig.~2b), the tails have the same chirality.  The asymptotic approach is slower in the unstable tail, due to the asymptotic behaviour $\eta \sim (c_1 + c_2 \zeta)e^{i k_u \zeta}$ as $\zeta \rightarrow +\infty$.  \color{black} The  parameters in (a) and (b) are the same as in Fig.~2a and Fig.~2b respectively.}
	\label{fig: asymptotics of profile}
\end{figure}

Supplementary Figure~\ref{fig: asymptotics of profile} shows the  profile of an emergent high-field domain wall obtained from numerical solution of the LLG eqution.  The transverse oscillations in the two tails of the profile can be regarded as entrained helical spin waves with complex wavenumbers $k_{s,u}$, 
where $s$ denotes the stable tail ($\zeta \rightarrow -\infty$) and $u$ the unstable tail ($\zeta \rightarrow +\infty$).  The imaginary parts of $k_{s,u}$ 
determine the spatial decay rate of the oscillations.  The wavenumbers extracted from the computed profiles coincide with the expressions
\begin{equation}
	\label{eq: kappa_us}
	k_s = \tfrac{i}{2} [r_s v - (8 + 2r_s^2 v^2)^{1/2} ], \ \ k_u =  \tfrac{i}{2} r_u v, \ \, 
\end{equation}
where $r_s = \alpha + i$ and $r_u = \alpha - i$; these are
obtained from the roots of Eq.~(6) with $v$ and $\omega$ given by Eq.~(15) ($k_s$ corresponds to the root of Eq.~(15) with negative imaginary part). It is straightforward to show that for $h_a$ greater than (resp.~less than) $3 - [2\alpha^2/(1+2\alpha^2)]$, the real parts of $k_s$ and $k_u$ have the same (resp.~opposite) signs; ie, the spin waves in the  tails have the same (resp.~opposite) chiralities -- cf Fig.~2.

\clearpage